\documentclass[prl,twocolumn,showpacs,amsmath,amssymb,superscriptaddress]{revtex4}
\usepackage{pifont,amssymb,epsf,graphicx}
\begin{document}
\title{In-gap Bound States Induced by Interstitial Fe Impurities in Iron-based Superconductors }

\author{Degang Zhang}
\affiliation{College of Physics and Electronic Engineering, Sichuan Normal University,
Chengdu 610101, China}
\affiliation{Institute of Solid State Physics, Sichuan Normal
University, Chengdu 610101, China}

\begin{abstract}

Based on a two-orbit four-band tight binding model, we investigate
the low-lying electronic states around the interstitial excess Fe ions in the
iron-based superconductors by using T-matrix approach. It is shown that
the local density of states at the interstitial Fe impurity (IFI) possesses
a strong resonance inside the gap, which seems to be insensitive to the doping and
the pairing symmetry in the Fe-Fe plane, while a single or two resonances
appear at the nearest neighboring (NN) Fe sites. The location and height of
the resonance peaks only depend on the hopping $t$ and the pairing parameter $\Delta_I$
between the IFI and the NN Fe sites. These in-gap resonances are originated in
the Andreev's bound states due to the quasiparticle tunneling through the IFI,
leading to the change of the magnitude of the superconducting order parameter.
When both $t$ and $\Delta_I$ are small, this robust zero-energy bound state
near the IFI is consistent with recent scanning tunneling
microscopy observations.

\end{abstract}


\maketitle

The discovery of different families of the iron-based superconductors
provided us a new platform to study high temperature superconductivity [1-6].
It has been reported that the superconducting transition temperature $T_c$
can reach as high as 55K [2]. Similar to the cuprate superconductors,
the iron-based superconductors also have a layer structure and
the superconductivity comes from the Cooper pairs in the Fe-Fe
plane by doping electrons or holes. However, different from the
oxygen ions at the Cu-Cu bonds in the cuprates, two As/Se/Te ions in
each unit cell locate just above (below) the center of one face
of the Fe square lattice and below (above) the center of the
neighboring face, respectively (see Fig. 1).
Such a combination of the Fe
ions and the As/Se/Te ions produces a more complex energy band structure.
Angle resolved photoemission spectroscopy (ARPES) experiments have
revealed that the iron-based superconductors usually possess
two hole Fermi surfaces around (0,0), i.e. $\Gamma$ point,
and two electron Fermi surfaces around $(\pi,\pi)$, i.e.
$M$ point [7-16]. We note that due to different dopings,
one or even two hole Fermi surfaces disappear in some families of
the iron-based superconductors. In order to explain these
Fermi surface characteristics, two-orbital, three-orbital, and five-orbital
tight binding models have been proposed [17-20].

\begin{figure}
\rotatebox[origin=c]{0}{\includegraphics[angle=0,
           height=2.1in]{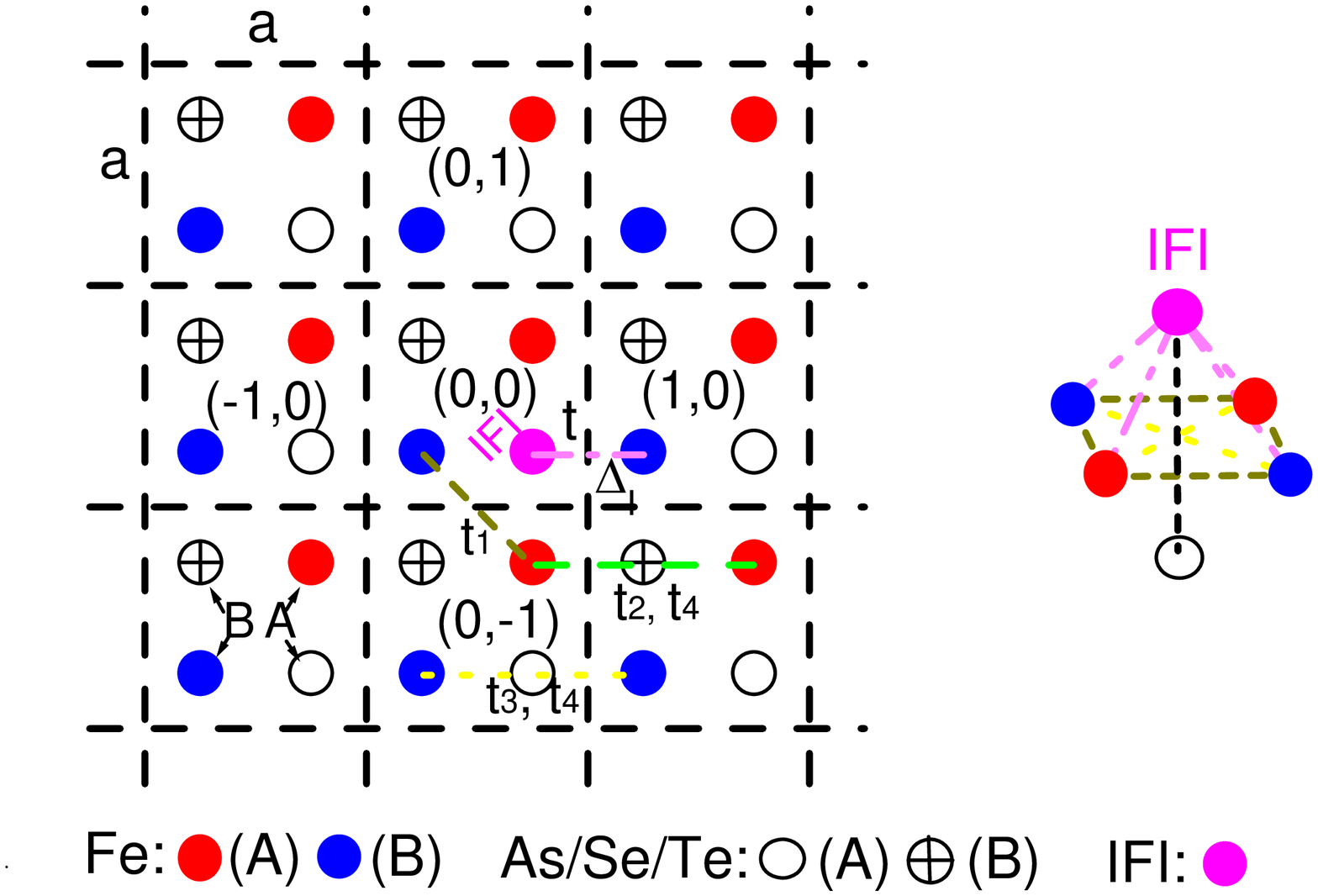}}
\caption{(Color online) Schematic lattice structure of the Fe-As/Se/Te
layers with each unit cell containing two
Fe (A and B) and two As/Se/Te (A and B) ions. The As/Se/Te ions A and B are
located just below and above the center of each face of the Fe
square lattice, respectively. Here, $t_1$ is the nearest neighboring (NN)
hopping between the same orbitals $d_{xz}$ or $d_{yz}$. $t_2$ and
$t_3$ are the next nearest neighboring hoppings between the same
orbitals mediated by the As/Se/Te ions B and A, respectively. $t_4$ is the
next nearest neighboring hopping between the different orbitals.
The IFI is situated at the symmetric position of  As/Se/Te ion about
the Fe-Fe plane. $t$ and $\Delta_I$ are the hopping and the pairing parameters
between the same orbitals of the IFI and the NN Fe sites,
respectively.}
\end{figure}

It is known that the impurity effects play a crucial role in exploring
superconductivity [21]. The strong zero-energy bound state (ZBS) at the Zn
ion and the fourfold symmetric electronic states around the single impurity,
observed by scanning tunneling microscopy (STM) [22], demonstrate that
the superconducting order parameter in the cuprates  has a d-wave symmetry.
The in-gap resonances induced by a nonmagnetic impurity was also used to
confirm the $s_{+-}$ pairing symmetry in the iron-based superconductors [18].
It is clear that these resonance peaks are originated in the
Andreev's bound states due to the quasiparticle scattering between
the Fermi surfaces or parts of the Fermi surface with opposite phases
of the superconducting order parameter.

Very recently, Yin {\it et al.} also used STM to study the electronic states
near the interstitial excess Fe ions in the iron-based superconductor
Fe(Te,Se) [23]. The as-grown Fe(Te,Se) single crystals  usually
contain a large amount of excess Fe ions situated randomly at
the interstitial sites, which are the symmetric points of the Te/Se
ions about the Fe-Fe plane (see Fig. 1).
They found that the differential
conductance at the interstitial Fe impurity (IFI) also has a strong ZBS.
The height of this resonance decays rapidly with the distance from the IFI
and vanishes at about $10\AA$. It is especially surprising that
the ZBS at the IFI is not affected by a magnetic field up to 8 Tesla.
Such a ZBS induced by out of the superconducting plane impurities was never
reported before, and its origin is unclear up to now.

In this work, we provide an explanation for the interesting observation of
the robust ZBS. Because the big family of the iron-based superconductors
has a similar energy band structure, here we employ the two-orbit four-band
tight binding model presented in Ref. [18] to explore the origin of the ZBS
near the IFI. We remember that this energy band structure
was constructed by starting two degenerate
orbitals $d_{xz}$ and $d_{yz}$ of per Fe ion and two Fe ions in each unit cell,
and fitted well the ARPES observations. Because the asymmetry
of the above and below As/Se/Te ions in the Fe-As/Se/Te surface layer was taken
into account, the model naturally explained several important STM observations,
e.g. in-gap impurity resonances [18,24], the bound state at negative energy
in the vortex core [25,26], the domain walls [27-30], etc.,
and especially repeated the phase diagram measured by
nuclear magnetic resonance and neutron scattering experiments [31-33].

It is obvious that the IFI is different from the nonmagnetic, magnetic, and Kondo
impurities in the superconducting plane. Because Fe ion is assumed to possess
two electron channels, the electrons in the Fe-Fe plane can tunnel through
the IFI. Now we know that superconductivity in the iron-based superconductors
originates from electron pairing between the next nearest neighboring Fe sites.
So only one electron of a Cooper pair can arrive at the IFI
in the superconducting state. Therefore,
the Hamiltonian describing the IFI in the iron-based superconductors
has the form

$$H=H_0+H_{\rm BCS}$$
$$-t\sum_{\alpha\sigma}\{c^+_{\alpha\sigma}[
c_{A\alpha,00\sigma}+c_{B\alpha,00\sigma}+
c_{A\alpha,0-1\sigma}+c_{B\alpha,10\sigma}]+{\rm h.c.}\}$$
$$+\Delta_I\sum_\alpha\{c_{\alpha\uparrow}^+[c_{A\alpha,00\downarrow}^+
+c_{A\alpha,0-1\downarrow}^++c_{B\alpha,00\downarrow}^+
+c_{B\alpha,10\downarrow}^+]$$
$$+c_{\alpha\downarrow}^+[c_{A\alpha,00\uparrow}^+
+c_{A\alpha,0-1\uparrow}^++c_{B\alpha,00\uparrow}^+
+c_{B\alpha,10\uparrow}^+]+{\rm h.c.}\},
\eqno{(1)}$$
where $H_0$ is the two-orbit four-band tight binding model proposed in
Ref. [18], $H_{\rm BCS}$ is the mean field BCS pairing Hamiltonian 
in the Fe-Fe plane, ${c}_{A(B)\alpha,ij\sigma }^{+}$
(${c}_{A(B)\alpha,ij\sigma}$) creates (destroys) an $\alpha$
electron with spin $\sigma$ (=$\uparrow$ or $\downarrow$) in the unit cell
$\{i, j\}$ of the sublattice A (B), $\alpha=0 (1)$  represents
the degenerate orbital $d_{xz}$ ($d_{yz}$), and
$t$ and $\Delta_I$ are the hopping and pairing parameters between 
the same orbitals of the IFI
and the nearest neighboring (NN) Fe sites, respectively.

We note that the sub-Hamiltonian $H_0+H_{\rm BCS}$ can be diagonalized
by suitable transformations. Introducing first the Fourier transformation
${c}_{A(B)\alpha,ij\sigma}=\frac{1}{\sqrt{N}}\sum_{\bf
k}{c}_{A(B)\alpha,{\bf k}\sigma}e^{i(k_x x_i+k_y y_j)}$ with $N$ the
number of unit cells and the canonical transformations
${c}_{A\alpha,{\bf k}\sigma}=\sum_{uv}(-1)^{\alpha v}
\frac{a_{u,{\bf k}}}{\Gamma_{u,{\bf k}}}\phi_{uv,{\bf k}\sigma}$ and
${c}_{B\alpha,{\bf k}\sigma}=\sum_{uv}(-1)^{\alpha v}
\frac{\epsilon_{T,{\bf k}}^*}{\Gamma_{u,{\bf k}}}\phi_{uv,{\bf k}\sigma}$
with $u,v=0$ and 1, $a_{u,{\bf
k}}=\frac{1}{2}(\epsilon_{A,{\bf k}}-\epsilon_{B,{\bf
k}})+(-1)^u\sqrt{\frac{1}{4}(\epsilon_{A,{\bf k}}-\epsilon_{B,{\bf
k}})^2+\epsilon_{T,{\bf k}}\epsilon_{T,{\bf k}}^*}$,
$\epsilon_{A,{\bf k}}=-2(t_2\cos k_x+t_3\cos k_y), \epsilon_{B,{\bf
k}}=-2(t_2\cos k_y+t_3\cos k_x)$, $\epsilon_{T,{\bf
k}}=-t_1[1+e^{ik_x}+e^{ik_y}+e^{i(k_x+k_y)}]$, and
$\Gamma_{u,{\bf k}}=\sqrt{2(a_{u,{\bf k}}^2+\epsilon_{T,{\bf
k}}\epsilon_{T,{\bf k}}^*)}$, and then taking the Bogoliubov
transformations $\phi_{uv{\bf k}\uparrow}=\sum_{\nu=0,1}
(-1)^\nu\xi_{uv{\bf k}\nu}\psi_{uv{\bf k}\nu}$ and
$\phi_{uv-{\bf k}\downarrow}^+=\sum_{\nu=0,1}
\xi_{uv{\bf k}1-\nu}\psi_{uv{\bf k}\nu}$
with constants $\xi_{uv{\bf k}\nu}$ to be determined below,
the total Hamiltonian $H$ finally becomes
$$H=\sum_{uv{\bf k}\nu}(-1)^\nu\Omega_{uv{\bf k}}
\psi_{uv{\bf k}\nu}^+\psi_{uv{\bf k}\nu}$$
$$+\frac{1}{\sqrt{N}}\sum_{uv\alpha{\bf k}\nu}({\cal M}_{uv{\bf k}\nu,\alpha}
c^+_{\alpha\uparrow}\psi_{uv{\bf k}\nu}+
{\cal N}_{uv{\bf k}\nu,\alpha}c^+_{\alpha\downarrow}
\psi_{uv{\bf k}\nu}^++{\rm h.c.}),\eqno{(2)}$$
where $\Omega_{uv{\bf k}}=\sqrt{(E_{uv{\bf k}}-\mu)^2+\Delta_{uv{\bf k}}^2}$,
$E_{uv{\bf k}}=\frac{1}{2}(\epsilon_{A,{\bf k}}+\epsilon_{B,{\bf
k}})-2(-1)^vt_4(\cos k_x+\cos k_y)
+(-1)^u\sqrt{\frac{1}{4}(\epsilon_{A,{\bf k}}-\epsilon_{B,{\bf
k}})^2+\epsilon_{T,{\bf k}}\epsilon_{T,{\bf k}}^*}$,
${\cal M}_{uv{\bf k}\nu,\alpha}=(-1)^{\alpha v}\tau_{u,{\bf k}}^*
[-(-1)^\nu t\xi_{uv{\bf k}\nu}+\Delta_I\xi_{uv{\bf k}1-\nu}]$,
${\cal N}_{uv{\bf k}\nu,\alpha}=(-1)^{\alpha v}\tau_{u,{\bf k}}
[-t\xi_{uv{\bf k}1-\nu}+(-1)^\nu\Delta_I\xi_{uv{\bf k}\nu}]$,
$\tau_{u,{\bf k}}=[(1+e^{ik_y})a_{u,{\bf k}}+
(1+e^{-ik_x})\epsilon_{T,{\bf k}}]/\Gamma_{u,{\bf k}}$,
$\xi_{uv{\bf k}0}\xi_{uv{\bf k}1}=\Delta_{uv{\bf k}}/(2\Omega_{uv{\bf k}})$,
and $\xi_{uv{\bf k}\nu}^2=\frac{1}{2}[1+(-1)^\nu (E_{uv{\bf k}}-\mu)/
\Omega_{uv{\bf k}}]$.
Here, $\mu$ is the chemical potential to be determined by doping and
$\Delta_{uv{\bf k}}$ is the superconducting order parameter in the
Fe-Fe plane.
We can see from the Hamiltonian (2) that the hopping $t$ may
form new electron-fermion pairs while the pairing parameter
$\Delta_I$ can also induce a hopping term.
In other words, new quasiparticles are excited if electrons
tunnel through the IFI. It is expected that such a novel mechanism
produces the ZBS observed by the STM [23].

\begin{figure}
\rotatebox[origin=c]{0}{\includegraphics[angle=0,
           height=1.2in]{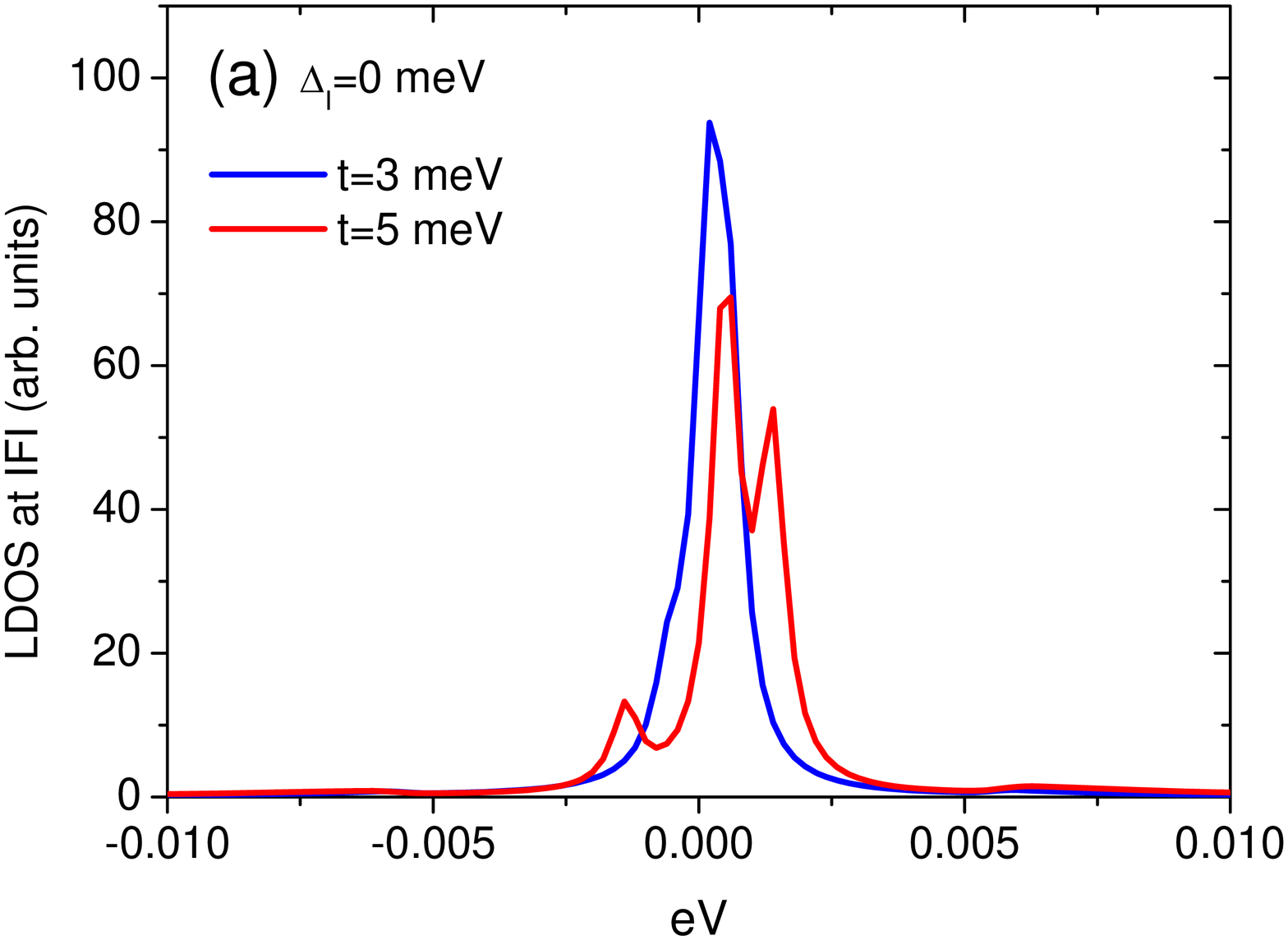}}
\rotatebox[origin=c]{0}{\includegraphics[angle=0,
           height=1.2in]{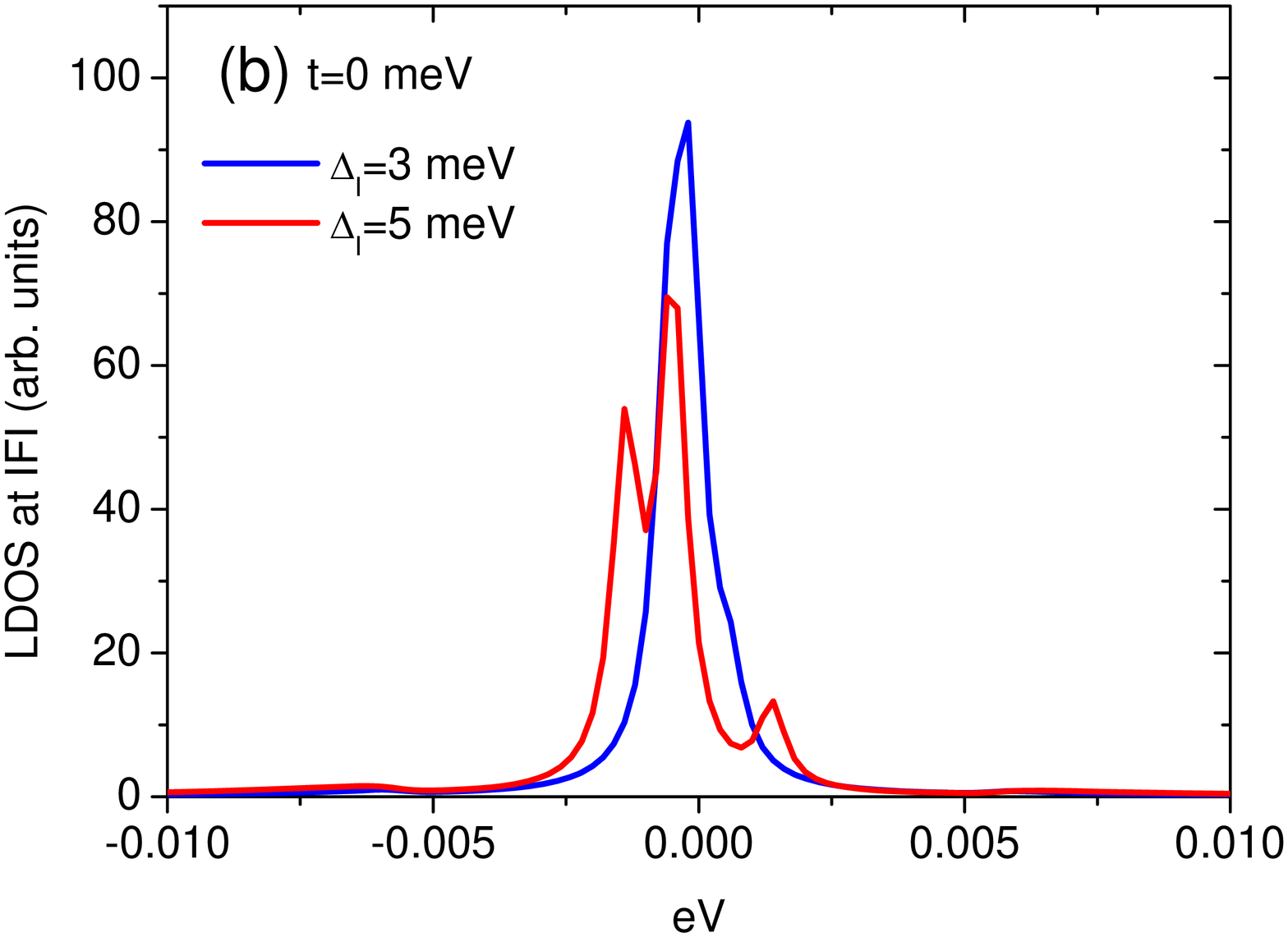}}
\rotatebox[origin=c]{0}{\includegraphics[angle=0,
           height=1.2in]{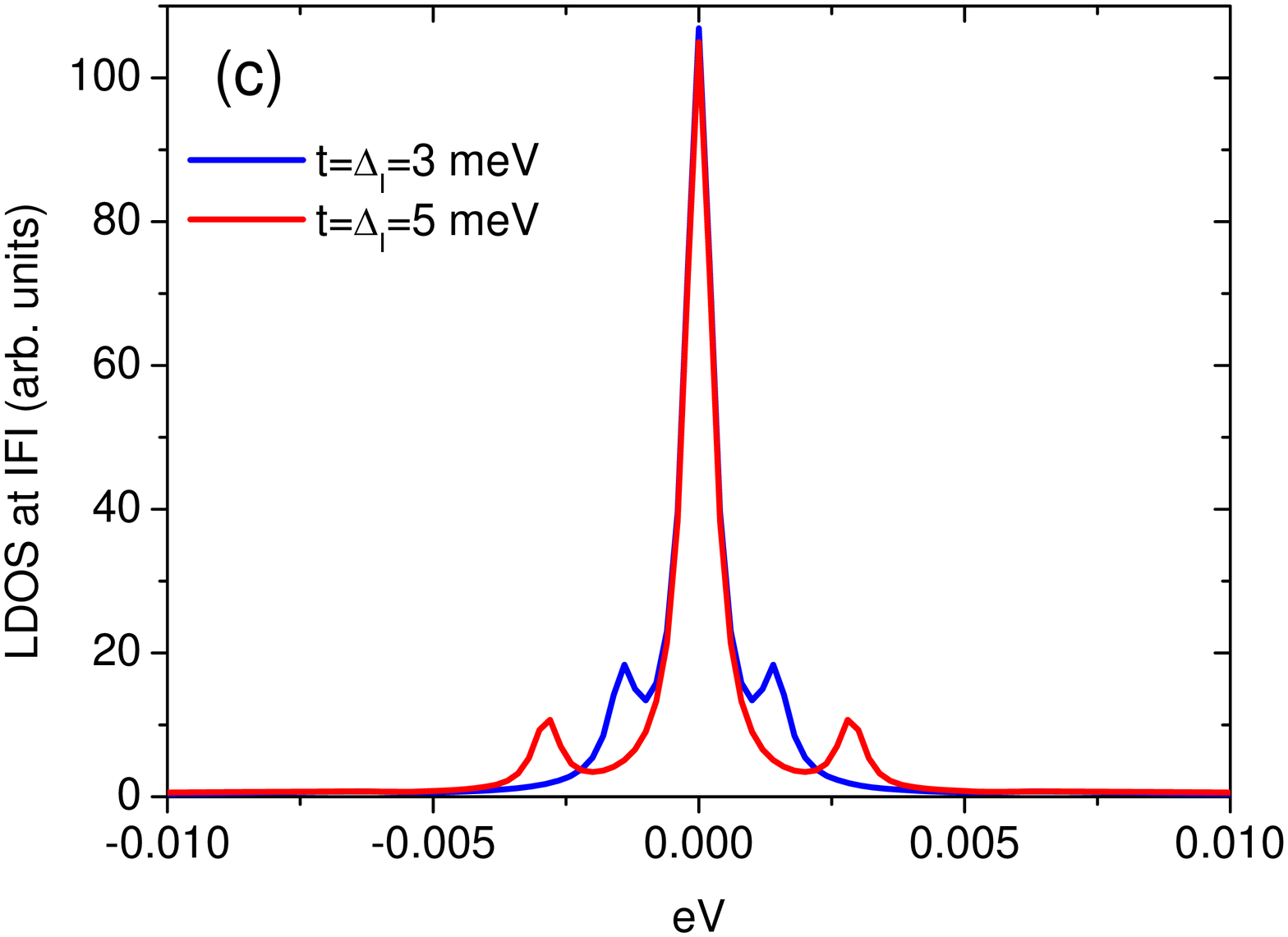}}
\rotatebox[origin=c]{0}{\includegraphics[angle=0,
           height=1.2in]{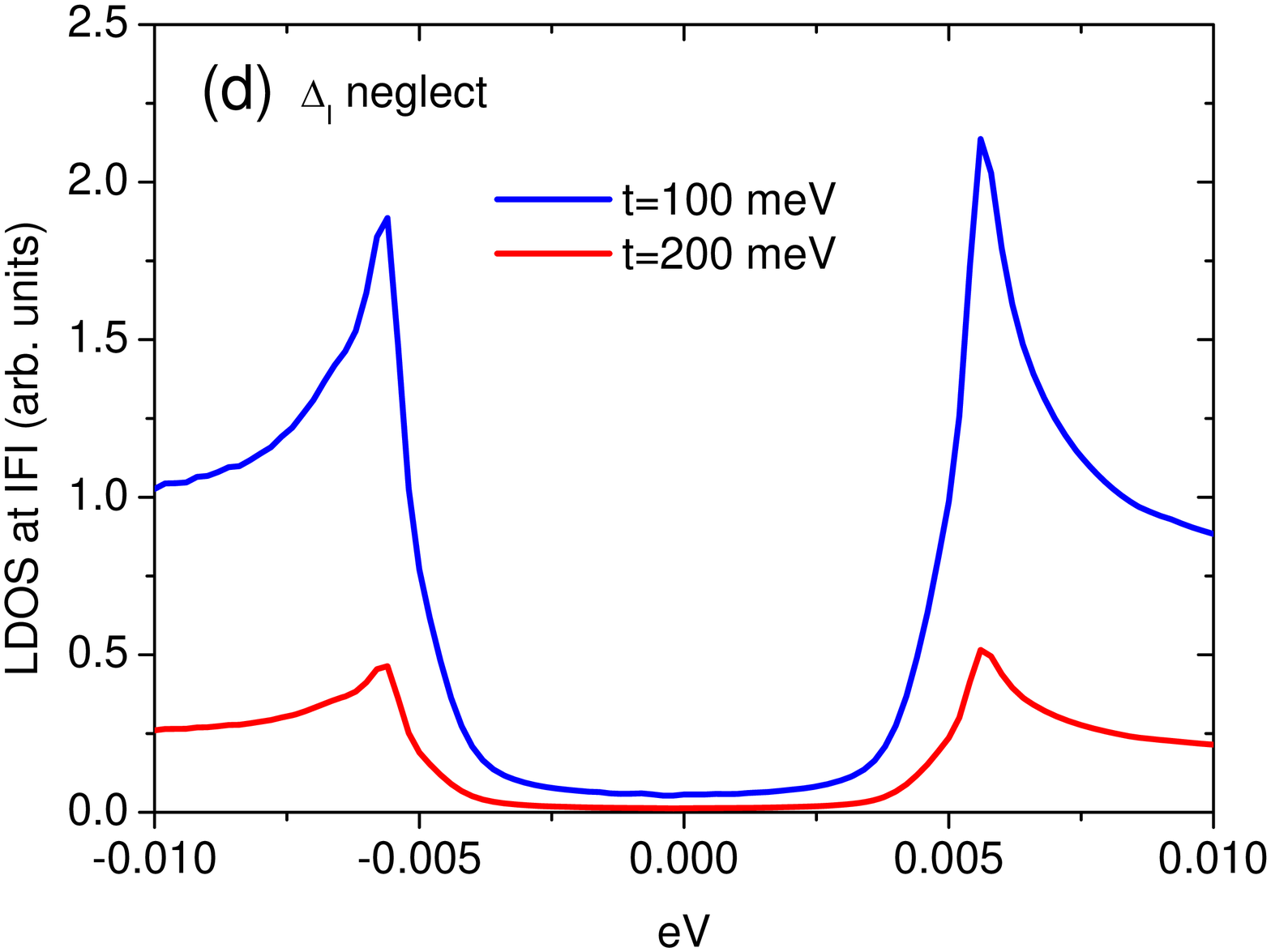}}

\caption {(Color online) The LDOS at the IFI  as a
function of the bias voltage eV under different $t$ and $\Delta_I$
for the $s_{+-}$ pairing symmetry $
\Delta_{uv{\rm \bf k}}=\frac{1}{2}\Delta_0(\cos k_x+\cos k_y)$ with
$\Delta_0=5.8$ meV  at optimal
electron doping ($\mu=-0.54$), where there are two Fermi surface sheets around
$\Gamma$ point.}
\end{figure}

The Hamiltonian $H$ in Eq. (2) has a quadratic form of the fermionic
operators and can be exactly solved by using the T-matrix approach [18,21].
The analytic expression of the LDOS near the IFI has been obtained
at different bias voltages. In our calculations, we have used the
energy band parameters: $t_1=0.5, t_2=0.2, t_3=-1.0$, and $t_4=0.02$.
We note that the iron-based superconductor Fe(Te,Se) in the STM
experiment [23] has a small gap of about 1.5 meV. In order to clearly
observe the change of the impurity resonance peaks with the bias
voltage, here we employ a large gap, i.e. 5.8 meV, in the
electron-doped BaFe$_{2-x}$Co$_x$As$_2$ [34,35], where the excess Fe
or Co ion is also regarded as an IFI.
In Fig. 2, we plot the LDOS at the IFI as a function of the bias voltage
$eV$ under different $t$ and $\Delta_I$ for the $s_{+-}$ pairing symmetry
at the chemical potential $\mu=-0.54$, where there are two Fermi surface
sheets around the center of the Brillouin zone, i.e. $\Gamma$ point.
Obviously, when $t$ or $\Delta_I$ is small,
i.e. $<\sim 3$ meV, the LDOS has a strong
resonance peak at zero energy, and the superconducting coherence peaks
do not show up. With increasing $t$ or $\Delta_I$, the sharp zero energy
resonance peak moves toward positive or negative energy, respectively,
and the superconducting coherence peaks appear at
the symmetric bias voltages in Fig. 2(a) and (b).
The height of the coherence peak at the positive energy
is higher (lower) than that at the negative energy for the parameter
$t$ ($\Delta_I$). We note that when $t<\sim 4$ meV and $\Delta_I=0$ meV,
where the coherence peak at the negative energy disappears, the LDOS at
the IFI has a single- or double-peak structure at the positive energies,
very similar to the STM observations in the black regions of the
superconducting BaFe$_{2-x}$Co$_x$As$_2$ sample [35].
When $t=\Delta_I=$3 and 5 meV, the LDOS possesses
the strong ZBS and the symmetric coherence peaks, shown in Fig. 2(c).
However, the superconducting energy gap at the
IFI is smaller that in the bulk, i.e. $\Delta_0=5.8$ meV, and becomes
larger if $t$ or $\Delta_I$ increases. When $t$ is large enough to
neglect $\Delta_I$, the superconducting energy gap at the
IFI is equal to $\Delta_0$, and there is no in-gap bound state in
Fig. 2(d).

\begin{figure}
\rotatebox[origin=c]{0}{\includegraphics[angle=0,
           height=1.2in]{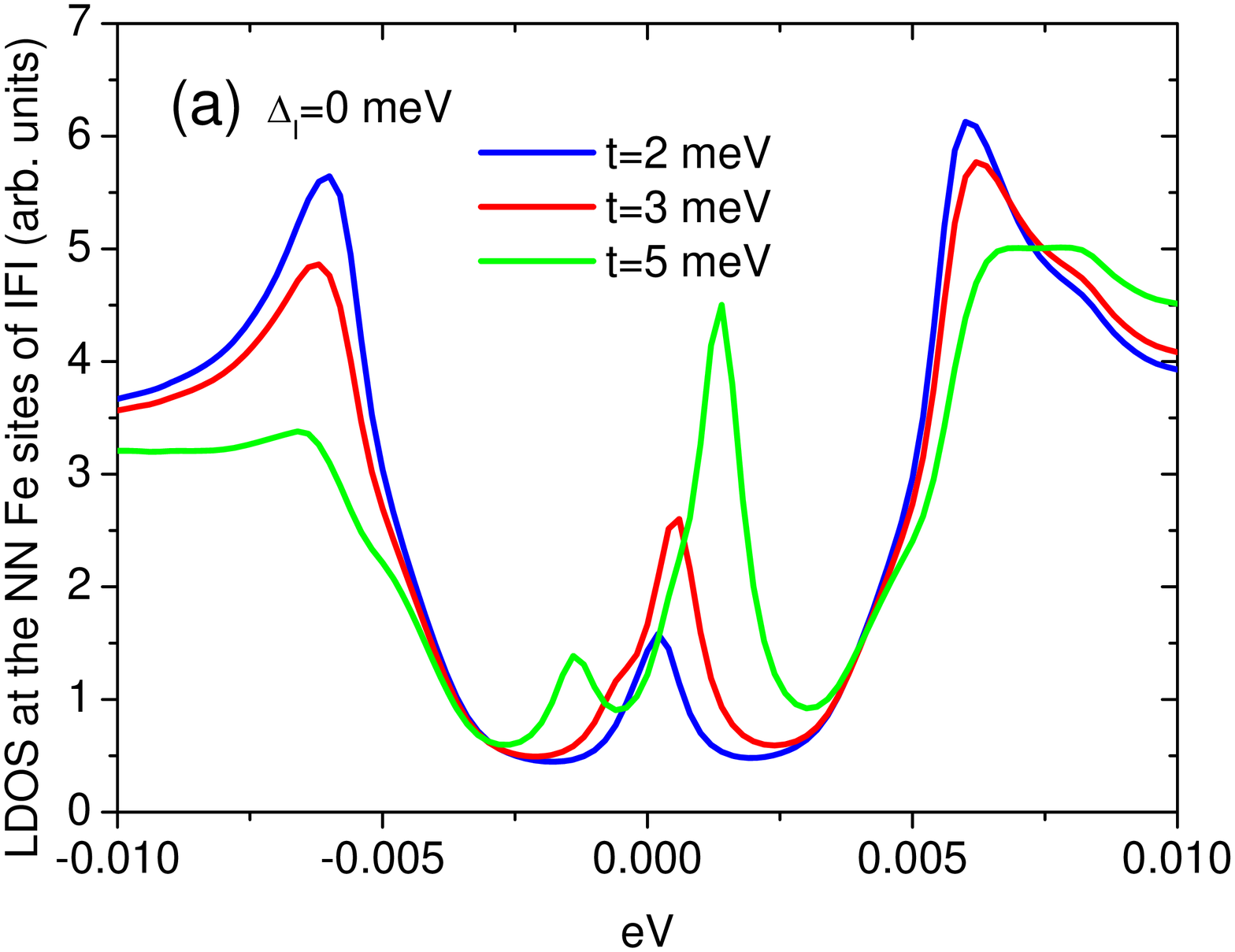}}
\rotatebox[origin=c]{0}{\includegraphics[angle=0,
           height=1.2in]{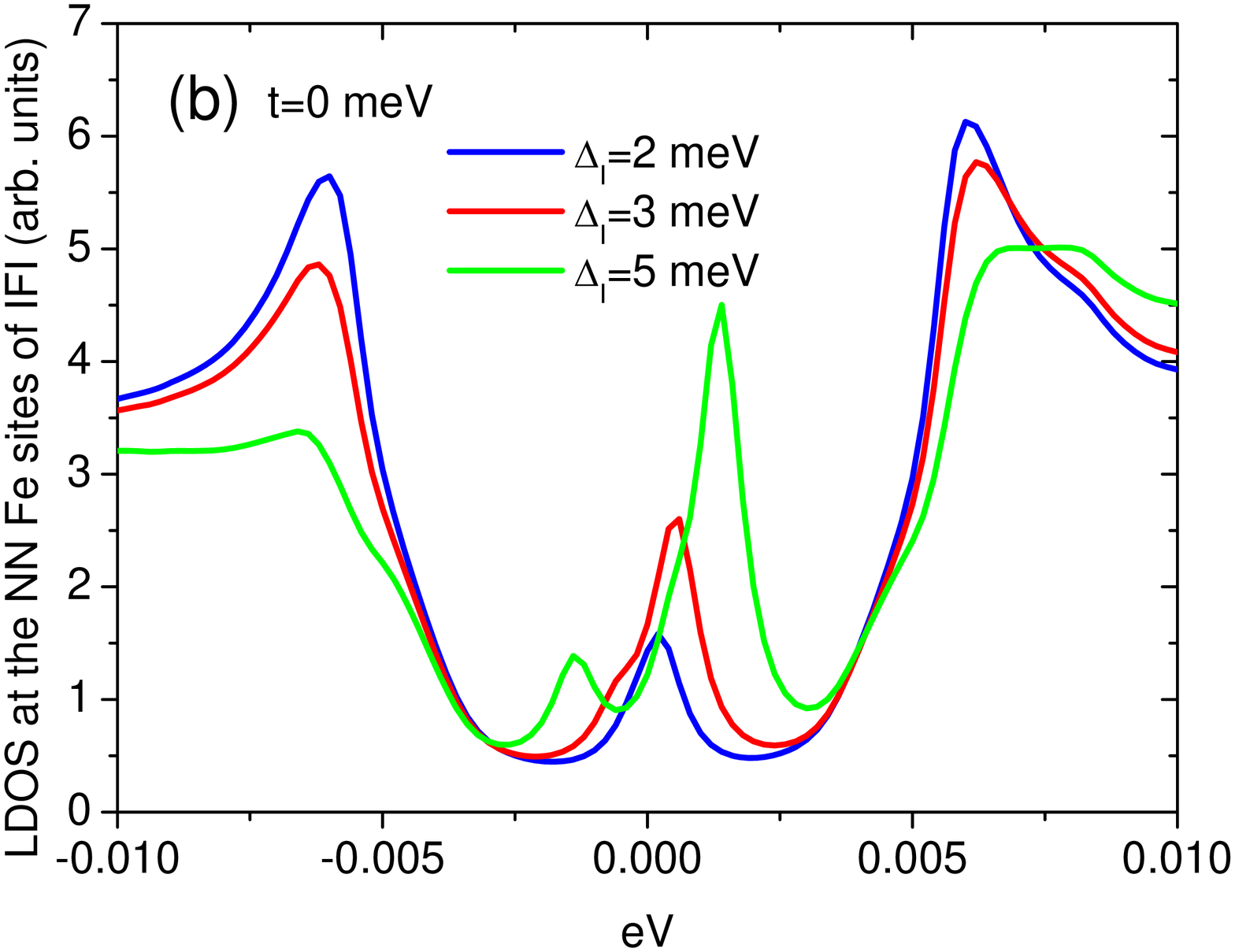}}
\rotatebox[origin=c]{0}{\includegraphics[angle=0,
           height=1.2in]{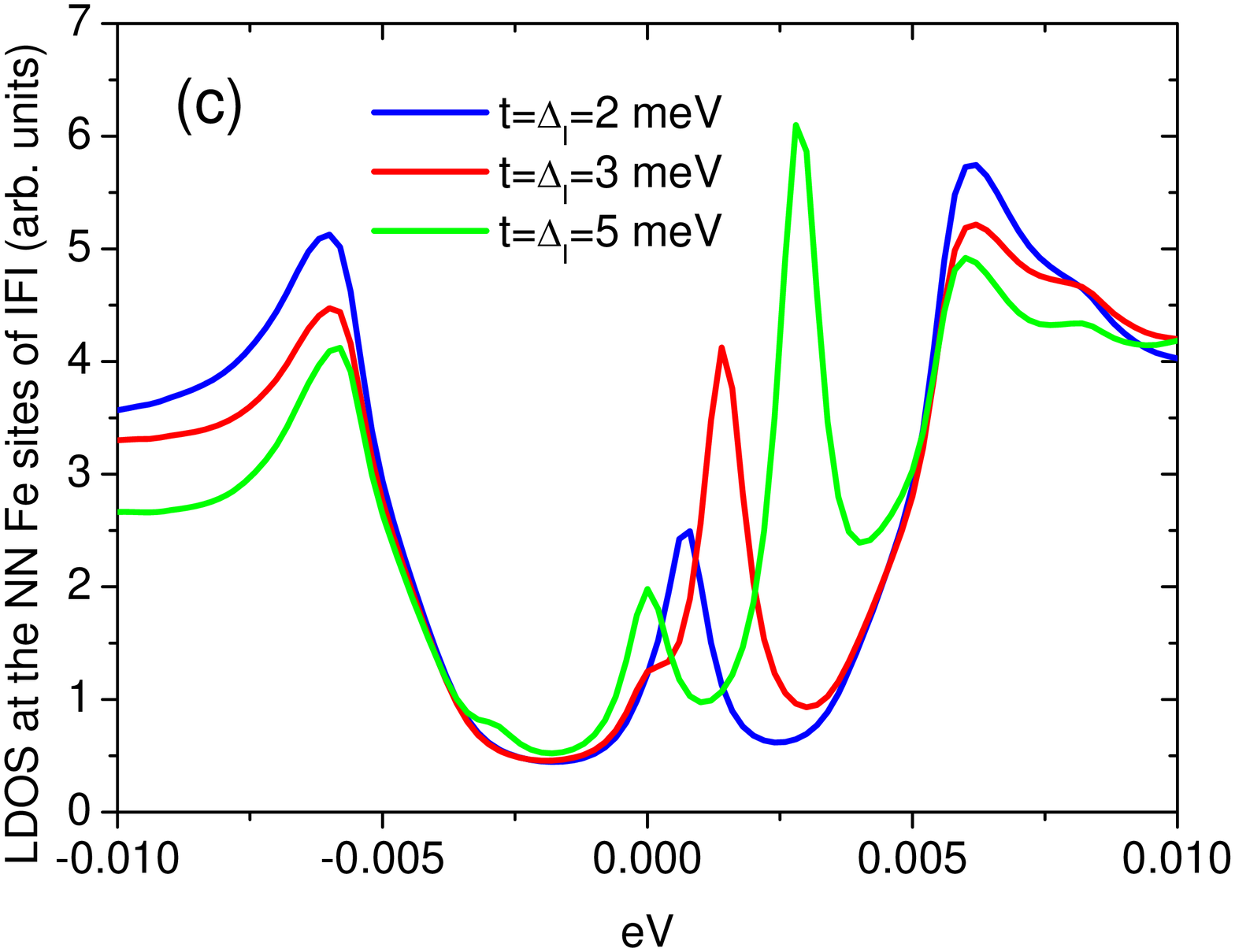}}
\rotatebox[origin=c]{0}{\includegraphics[angle=0,
           height=1.2in]{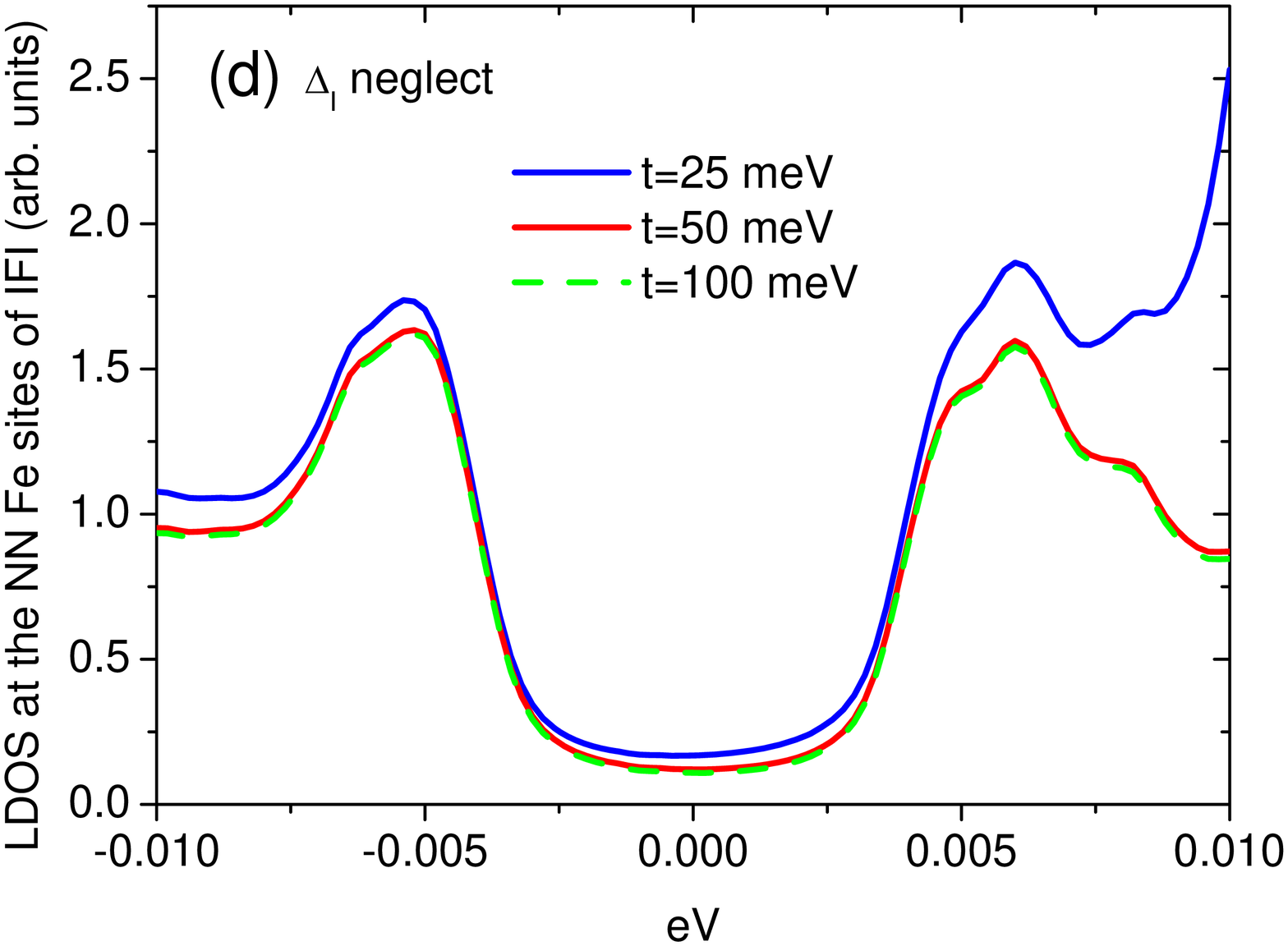}}

\caption {(Color online) The LDOS at the nearest neighboring Fe sites as a
function of the bias voltage eV under different $t$ and $\Delta_I$
for the $s_{+-}$ pairing symmetry $
\Delta_{uv{\rm \bf k}}=\frac{1}{2}\Delta_0(\cos k_x+\cos k_y)$ with
$\Delta_0=5.8$ meV  at optimal
electron doping ($\mu=-0.54$), where there are two Fermi surface sheets around
$\Gamma$ point.}
\end{figure}

Fig. 3 shows the LDOS at the NN Fe sites of the IFI for the
$s_{+-}$ pairing symmetry under $\mu=-0.54$. When $t (\Delta_I)\sim 2$ meV
and $\Delta_I (t)=0$ meV, there is a ZBS appeared in the LDOS
(see Fig. 3(a) and (b)). With increasing $t$ or $\Delta_I$,
the single resonance splits into two peaks at symmetric energies.
The resonance peak at positive energy has higher height than that at negative
energy. It is interesting that the curves of the LDOS in Fig. 3(a)
are same with those in Fig. 3(b) if $t$ and $\Delta_I$ are replaced each other.
In other words, $t$ and $\Delta_I$ play the same role for the LDOS
at the NN Fe sites of the IFI. This is the result that the Cooper pairs
in the Fe-Fe plane tunnel through the IFI.
We note that when $t=\Delta_I$, there is always a ZBS.
If $t=\Delta_I$ is large,  a strong resonance peak shows up
at positive bias voltages in Fig. 3(c). However, for $t>>\Delta_I$,
the in-gap resonances move outside of the superconducting
energy gap (see Fig. 3(d)).

\begin{figure}
\rotatebox[origin=c]{0}{\includegraphics[angle=0,
           height=1.2in]{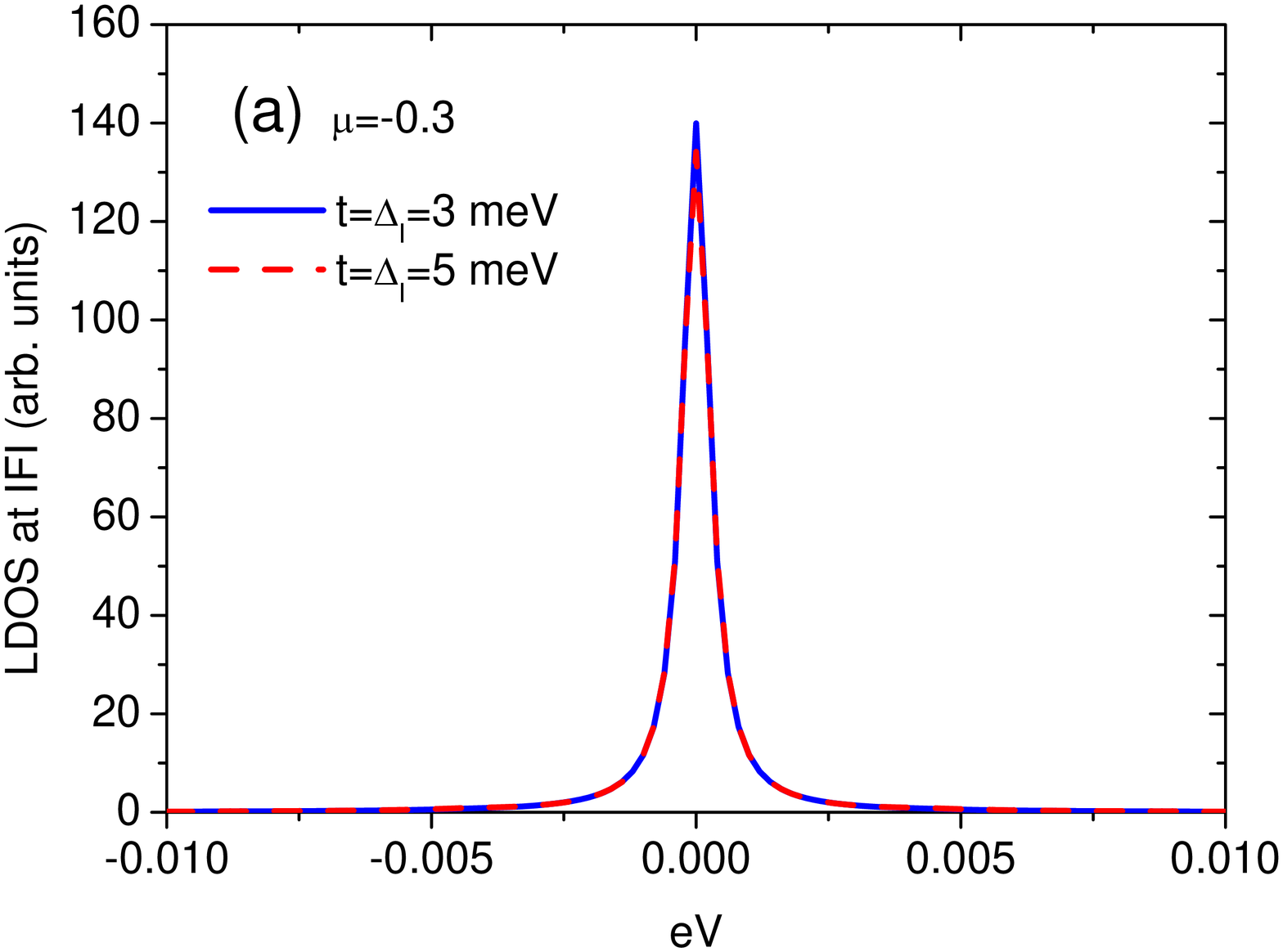}}
\rotatebox[origin=c]{0}{\includegraphics[angle=0,
           height=1.2in]{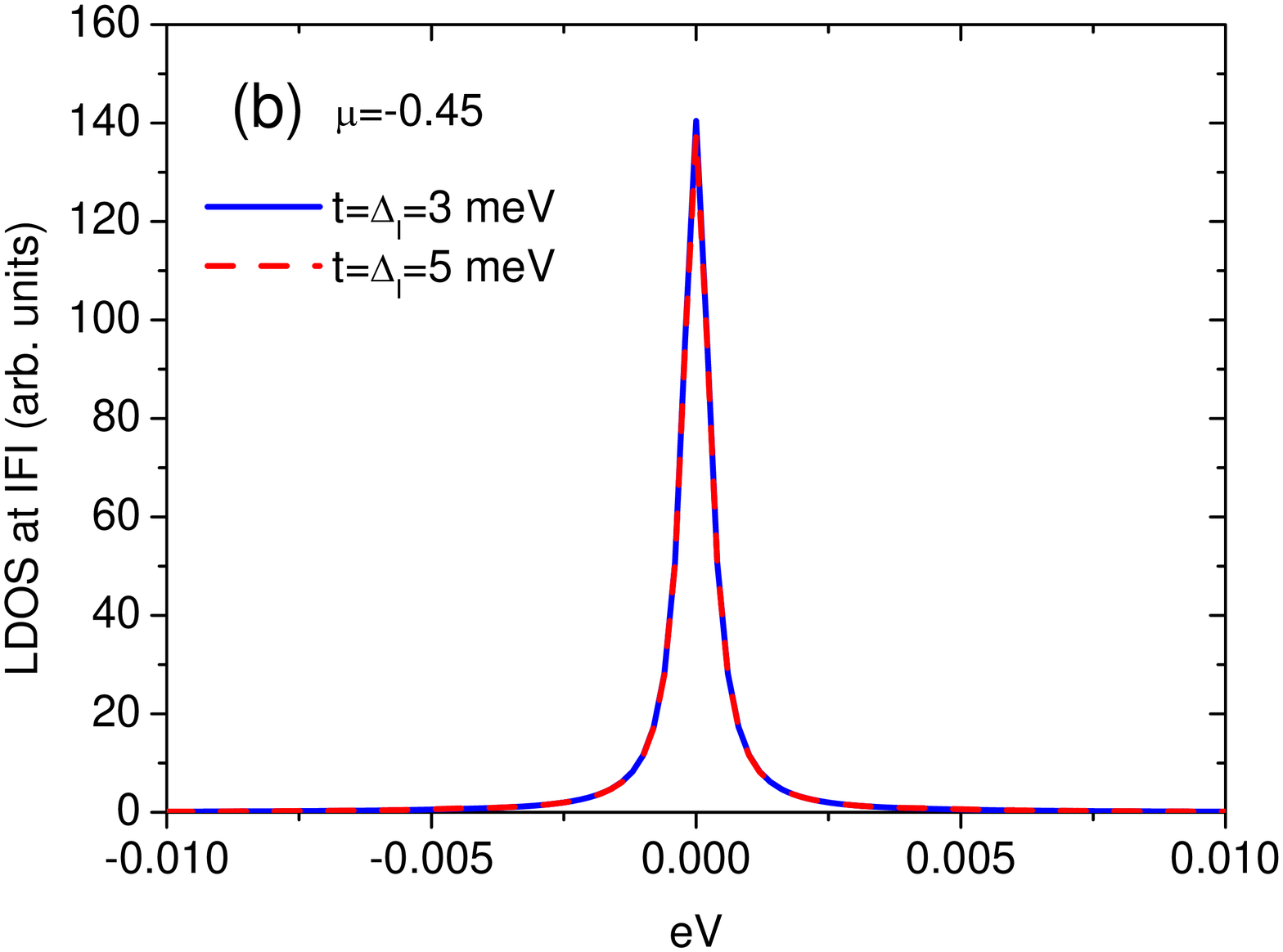}}
\rotatebox[origin=c]{0}{\includegraphics[angle=0,
           height=1.2in]{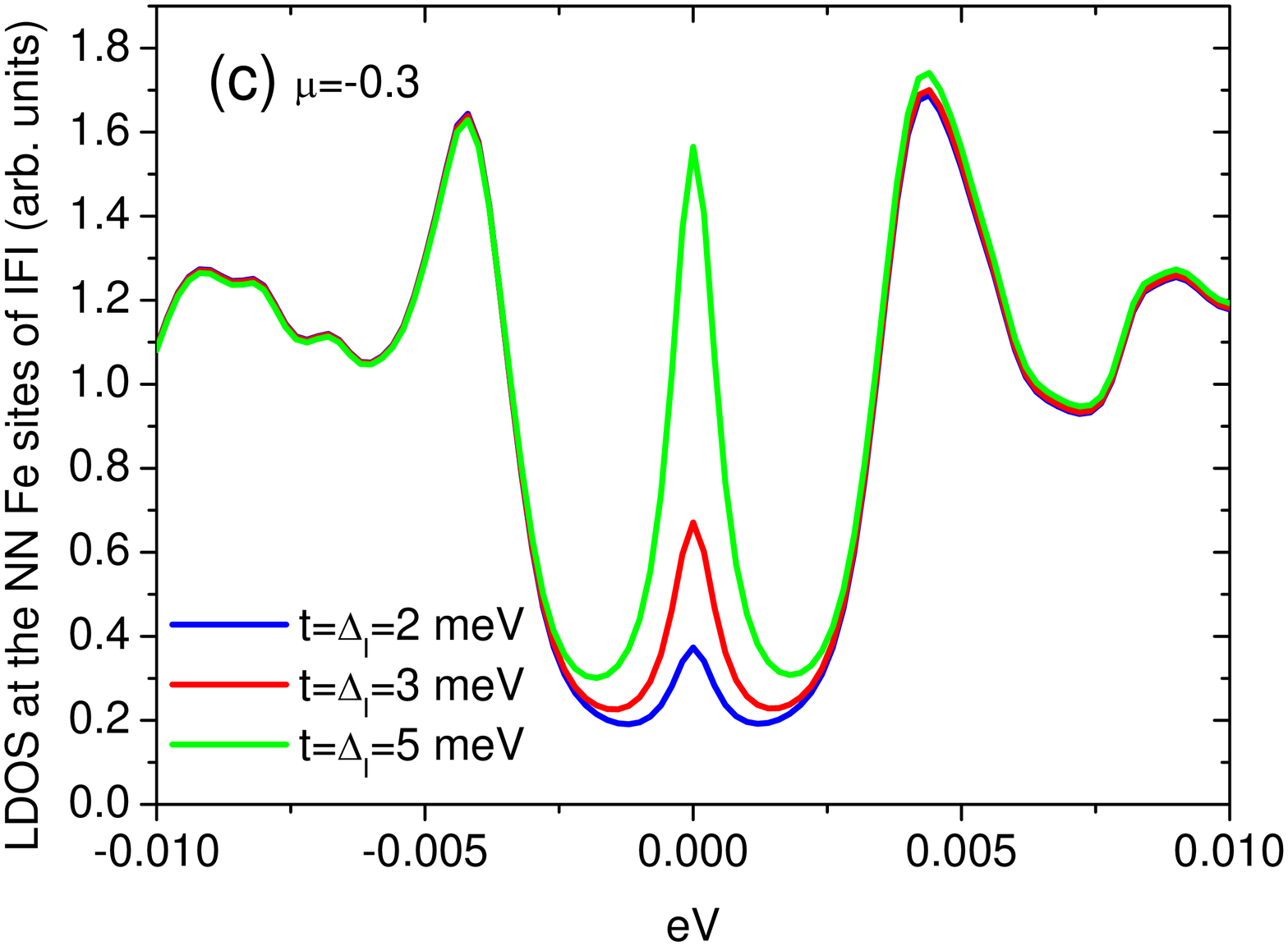}}
\rotatebox[origin=c]{0}{\includegraphics[angle=0,
           height=1.2in]{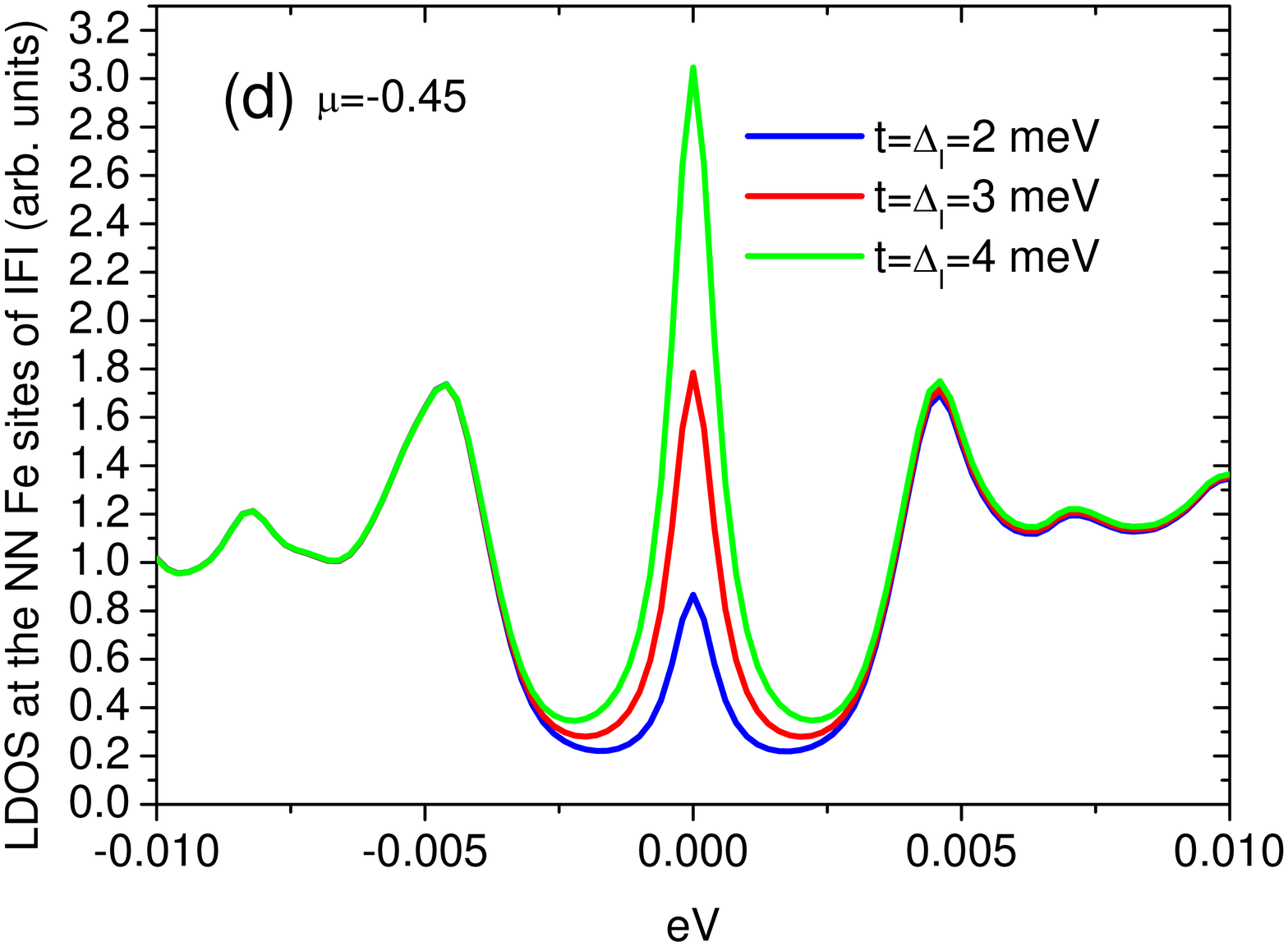}}

\caption {(Color online) The LDOS at the IFI in (a) and (b) and
the nearest neighboring Fe sites in (c) and (d) as a
function of the bias voltage eV under $\mu=-0.3$, where there is no
Fermi surface sheet around $\Gamma$ point, and $\mu=-0.45$, where there is one
Fermi surface sheet around $\Gamma$ point,
for the $s_{+-}$ pairing symmetry $
\Delta_{uv{\rm \bf k}}=\frac{1}{2}\Delta_0(\cos k_x+\cos k_y)$ with
$\Delta_0=5.8$ meV.}
\end{figure}

In order to understand the variation of the in-gap bound states
with the doping, we plot the LDOS curves near the IFI for the $s_{+-}$
pairing symmetry under $\mu=$-0.3 and -0.45, where there is no
Fermi surface sheet and one Fermi surface sheet around $\Gamma$ point,
respectively, in Fig. 4. Obviously, in both cases, the LDOS at the IFI
in Fig. 4(a) and (b) has a
sharp zero energy resonance, and the superconducting coherence peaks
are completely suppressed when $t=\Delta_I$. Meanwhile, there is only a
ZBS at the NN Fe sites of the IFI in Fig. 4(c) and (d).
The larger the parameter $t=\Delta_I$, the higher the zero energy
resonance peak.

\begin{figure}
\rotatebox[origin=c]{0}{\includegraphics[angle=0,
           height=1.2in]{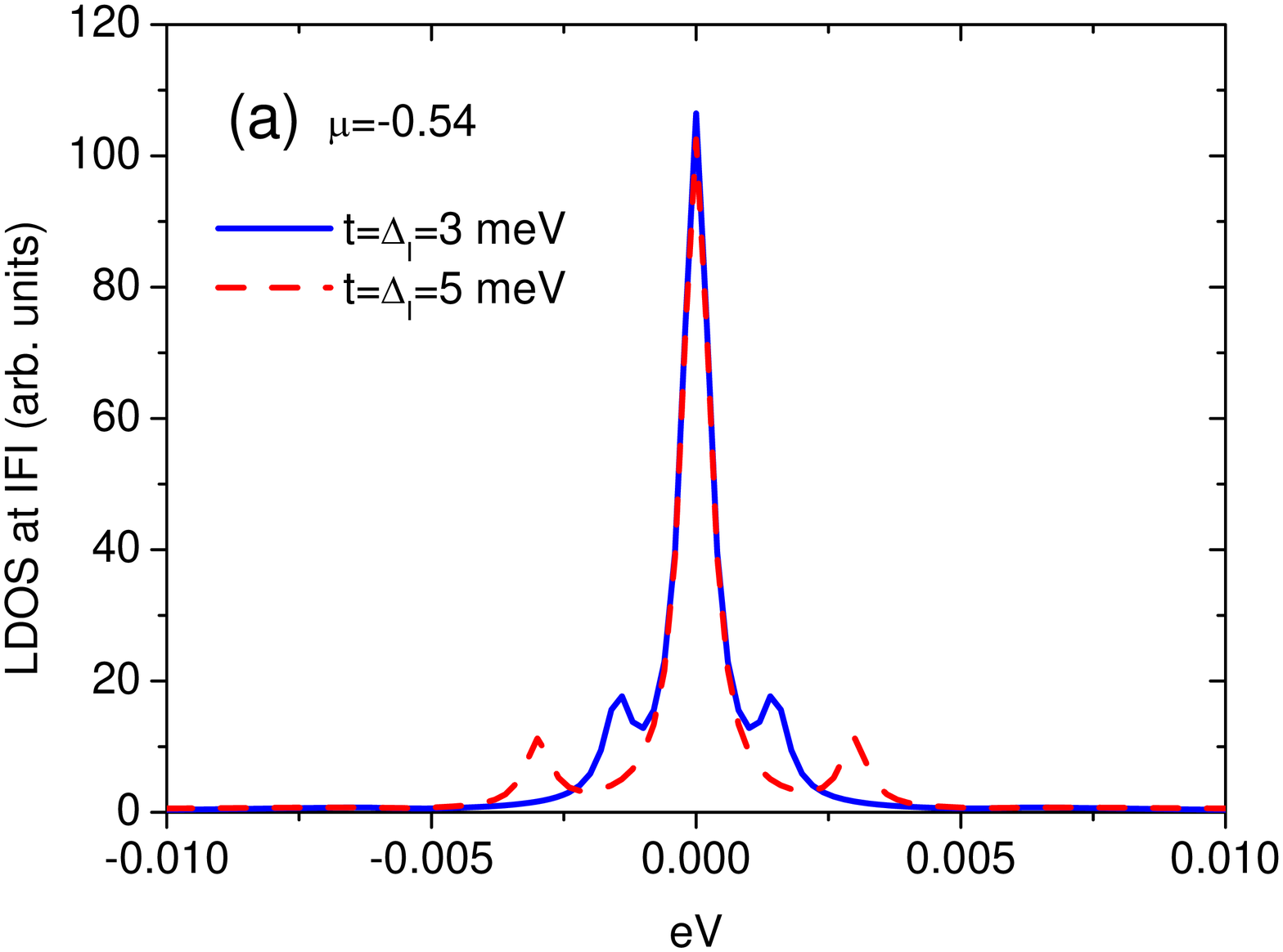}}
\rotatebox[origin=c]{0}{\includegraphics[angle=0,
           height=1.2in]{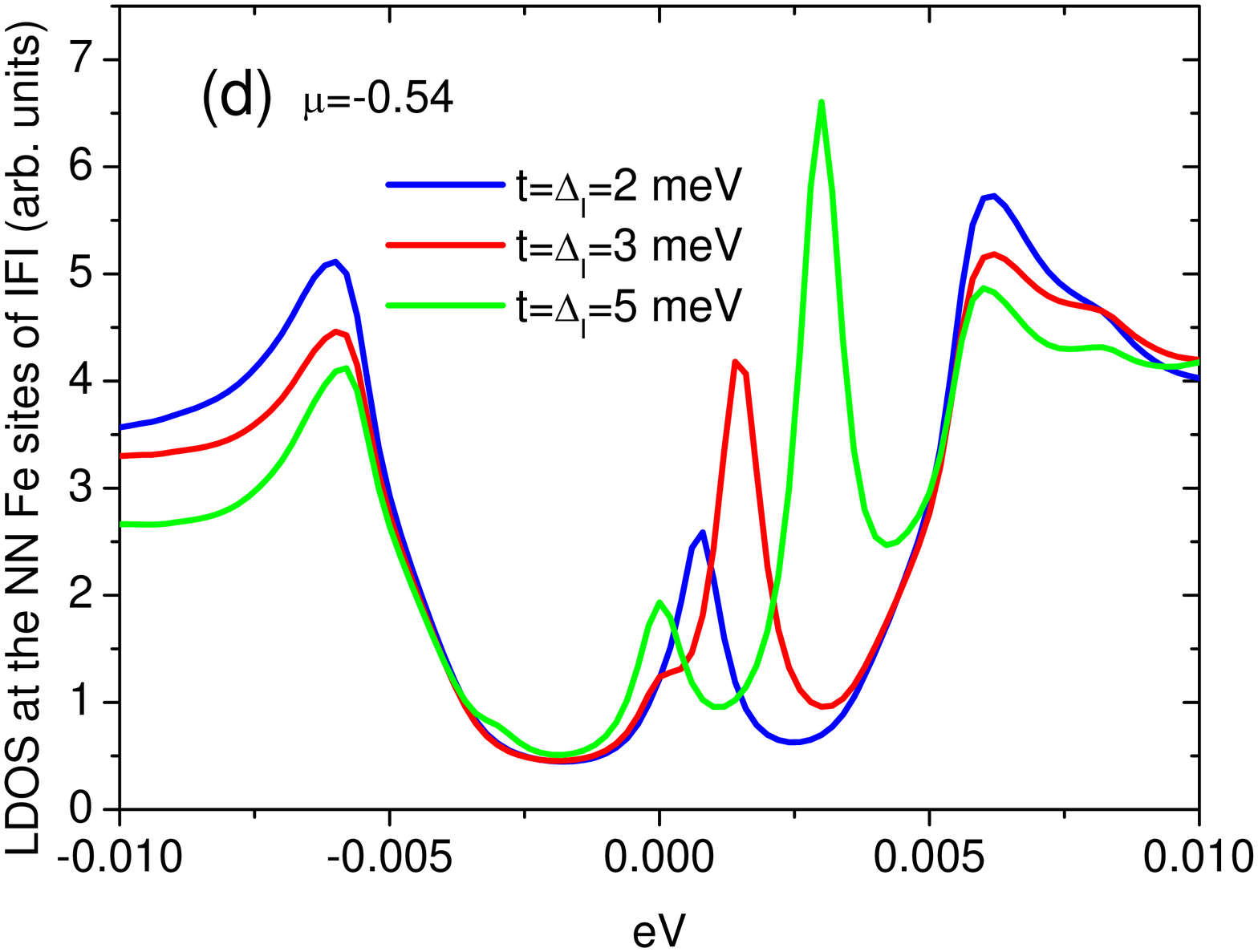}}
\rotatebox[origin=c]{0}{\includegraphics[angle=0,
           height=1.2in]{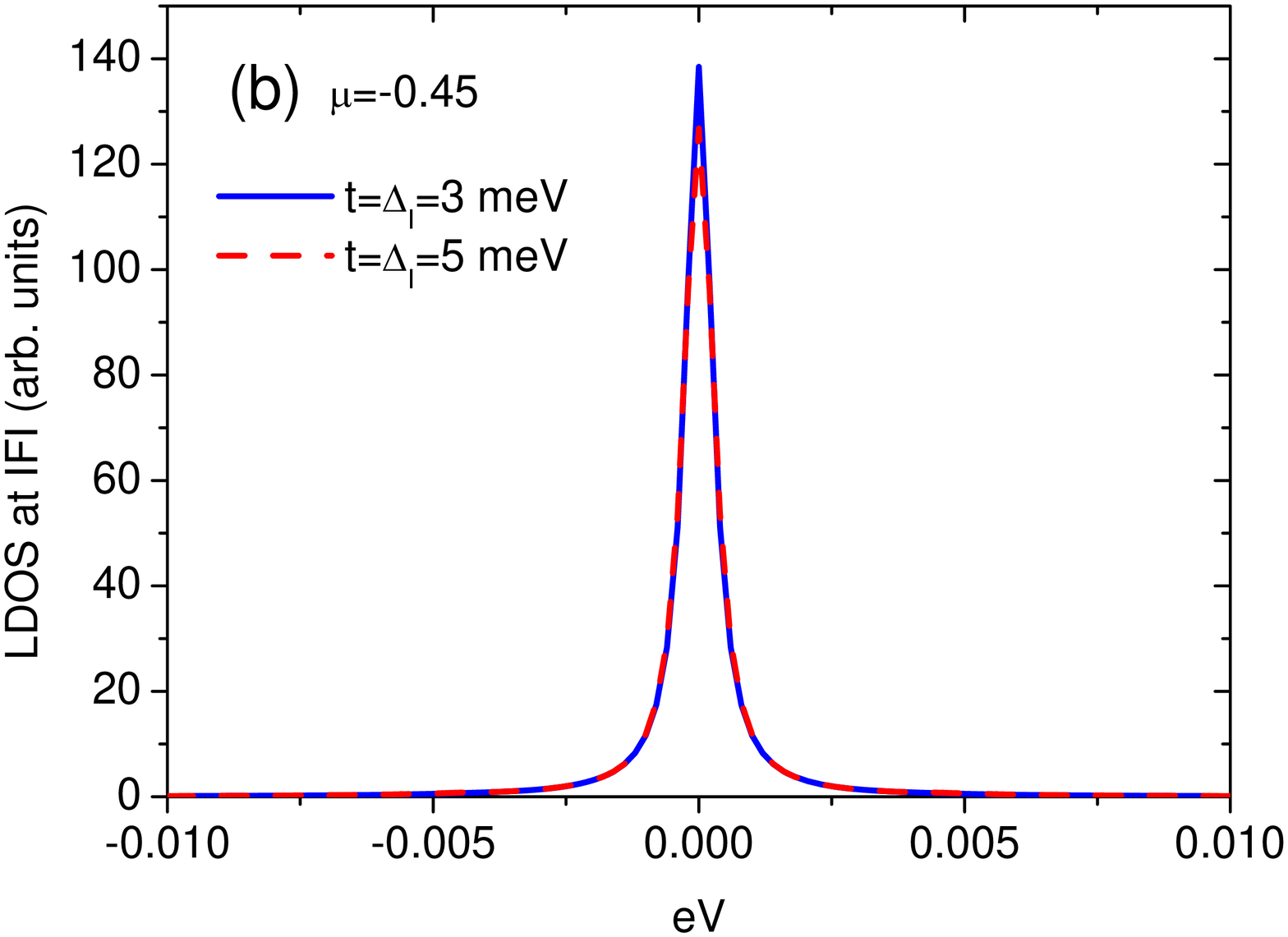}}
\rotatebox[origin=c]{0}{\includegraphics[angle=0,
           height=1.2in]{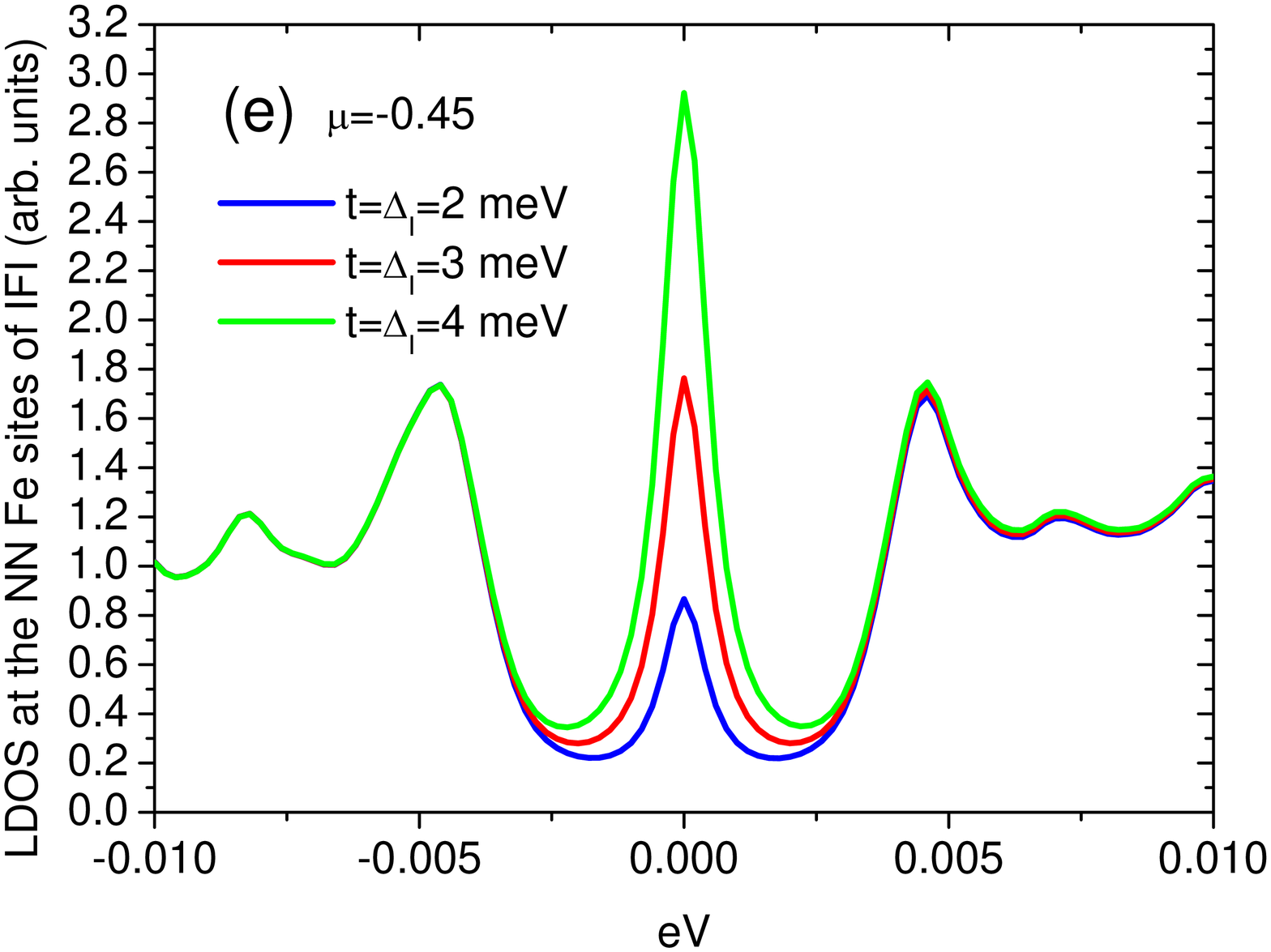}}
\rotatebox[origin=c]{0}{\includegraphics[angle=0,
           height=1.2in]{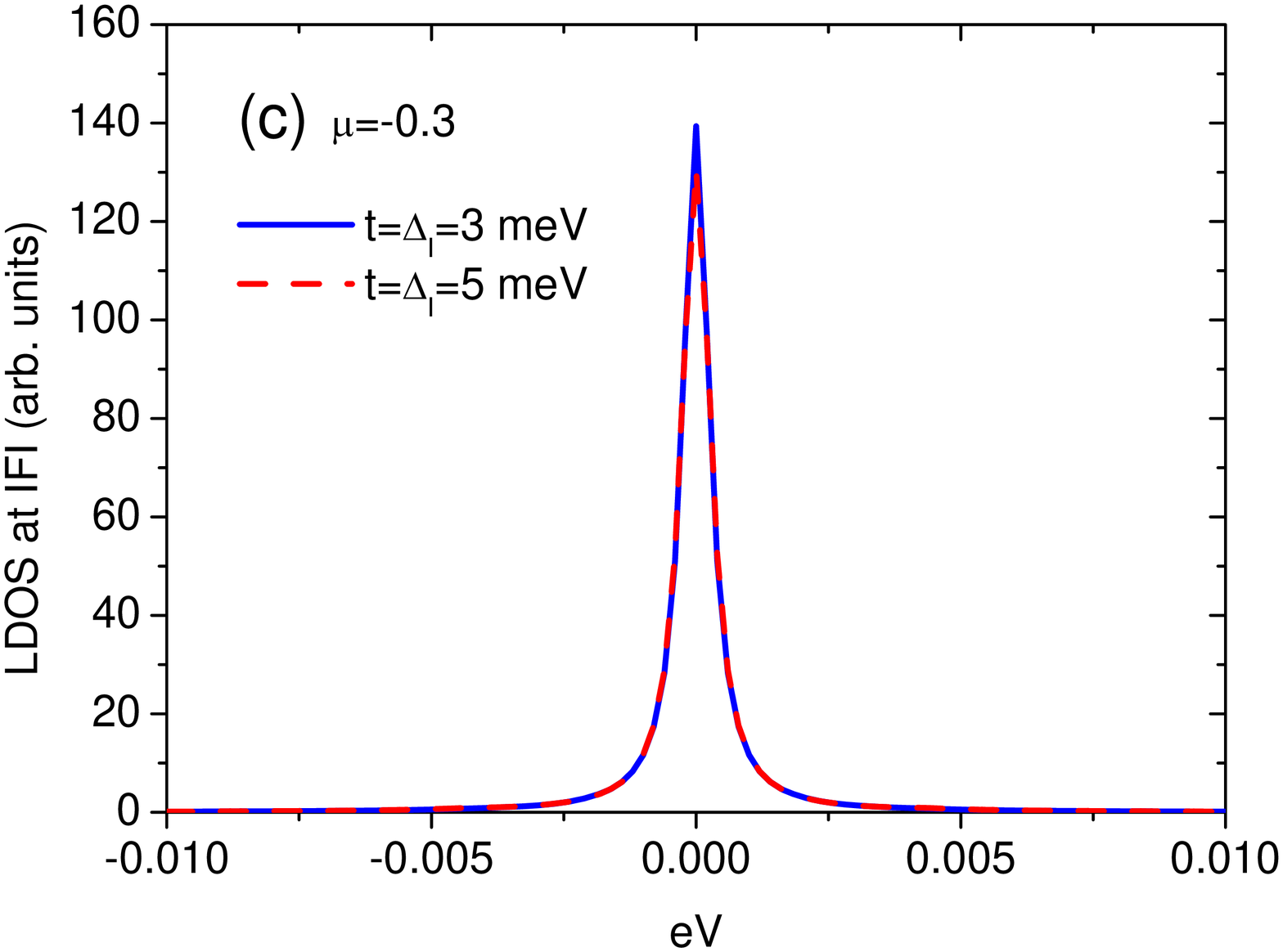}}
\rotatebox[origin=c]{0}{\includegraphics[angle=0,
           height=1.2in]{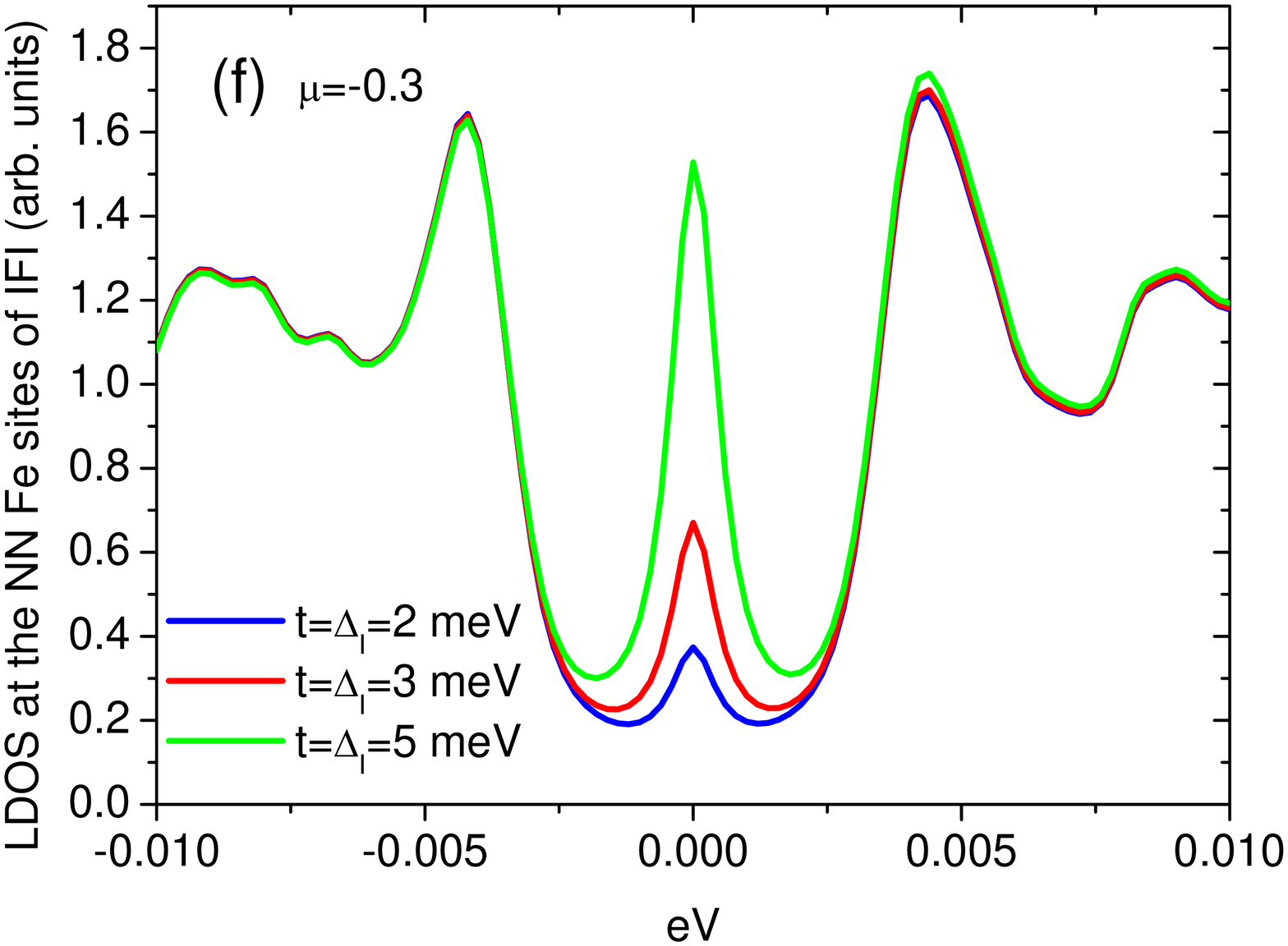}}

\caption {(Color online) The LDOS at the IFI in (a)-(c) and
the nearest neighboring Fe sites in (d)-(f)  as a
function of the bias voltage eV under different $t$ and $\Delta_I$
for the $s_{++}$ pairing symmetry $
\Delta_{uv{\rm \bf k}}=\frac{1}{2}\Delta_0|(\cos k_x+\cos k_y)|$ with
$\Delta_0=5.8$ meV  at different electron dopings.}
\end{figure}

Now we investigate the low-lying electronic states induced by the IFI
under different pairing symmetry in the bulk. In Fig. 5, we present the
LDOS near the IFI at different physical parameters for the $s_{++}$ pairing
symmetry. After comparing Fig. 5 with Fig. 2(c), Fig. 3(c), and
Fig. 4, we find that the LODS curves for the $s_{+-}$
pairing symmetry are identical with those for the $s_{++}$
pairing symmetry  in the presence of the same
physical parameters. This means that the in-gap resonance peaks induced
by the IFI is irrelative to the symmetry of the superconducting
order parameter. Therefore, this out of the superconducting plane IFI
is apparently different from the Zn impurity in cuprates [21,22]
and the nonmagnetic impurity in the iron-based superconductors [18,24].

In summary, we have explored the impact of the IFI on the electronic
states in the iron-based superconductors. This IFI is dealt as two electron
channels with the orbital index $\alpha$, which can also break the Cooper
pairs when electrons tunnel through it. It is undoubted that
re-forming new Cooper pairs with different magnitudes of the
superconducting order parameter produces the in-gap resonances,
which are determined by $t$ and $\Delta_I$ and
are independent of the pairing symmetry in the bulk. This implies that
the resonance peaks induced by the IFI are not affected by
a magnetic field. Obviously,
the origin of these in-gap bound states is due to the Andreev
reflection with the change of the magnitude rather than the phase of
the superconducting order parameter, which is never reported
in the superconductivity literature.
We note that when both $t$ and $\Delta_I$ are small,
the LDOS near the IFI always has a single resonance peak 
at zero bias voltage, which is also irrelative to the doping.
Such a robust ZBS is consistent with recent STM observations
in the iron-based superconductor Fe(Te,Se) [23].

The author would like to thank Jiaxin Yin and Ang Li for useful discussions.
This work was supported by the Sichuan Normal University.


\end{document}